\documentclass[sigconf]{acmart}
\usepackage{graphicx}
\usepackage{textcomp}
\usepackage{xcolor}
\usepackage{enumitem}
\usepackage{tabularx}
\usepackage{booktabs}
\usepackage{multirow}
\usepackage{svg}
\usepackage{multicol}
\usepackage{adjustbox}
\usepackage{url}

\AtBeginDocument{%
  }

\setcopyright{acmlicensed}
\copyrightyear{2024}
\acmYear{2024}
\acmDOI{XXXXXXX.XXXXXXX}

\acmConference[CF'24]{21st ACM International Conference on Computing Frontiers}{May 7--9, 2024}{Ischia, Italy}
\acmISBN{978-1-4503-XXXX-X/18/06}

\begin{document}

\title{Assessing the Performance of OpenTitan as Cryptographic Accelerator in Secure Open-Hardware System-on-Chips}

\author{Emanuele Parisi}
\email{emanuele.parisi@unibo.it}
\orcid{0000-0001-6607-7367}
\affiliation{
  \institution{University of Bologna}
  \streetaddress{Viale del Risorgimento, 2}
  \city{Bologna}
  \country{Italy}
}
\author{Alberto Musa}
\email{alberto.musa@unibo.it}
\orcid{0009-0003-1912-3801}
\affiliation{
  \institution{University of Bologna}
  \streetaddress{Viale del Risorgimento, 2}
  \city{Bologna}
  \country{Italy}
}
\author{Maicol Ciani}
\email{maicol.ciani@unibo.it}
\orcid{0009-0003-7861-9129}
\affiliation{
  \institution{University of Bologna}
  \streetaddress{Viale del Risorgimento, 2}
  \city{Bologna}
  \country{Italy}
}
\author{Francesco Barchi}
\email{francesco.barchi@unibo.it}
\orcid{0000-0001-5155-6883}
\affiliation{
  \institution{University of Bologna}
  \streetaddress{Viale del Risorgimento, 2}
  \city{Bologna}
  \country{Italy}
}
\author{Davide Rossi}
\email{davide.rossi@unibo.it}
\orcid{0000-0002-0651-5393}
\affiliation{
  \institution{University of Bologna}
  \streetaddress{Viale del Risorgimento, 2}
  \city{Bologna}
  \country{Italy}
}
\author{Andrea Bartolini}
\email{a.bartolini@unibo.it}
\orcid{0000-0002-1148-2450}
\affiliation{
  \institution{University of Bologna}
  \streetaddress{Viale del Risorgimento, 2}
  \city{Bologna}
  \country{Italy}
}
\author{Andrea Acquaviva}
\email{andrea.acquaviva@unibo.it}
\orcid{0000-0002-7323-759X}
\affiliation{
  \institution{University of Bologna}
  \streetaddress{Viale del Risorgimento, 2}
  \city{Bologna}
  \country{Italy}
}

\renewcommand{\shortauthors}{Parisi et al.}

\begin{abstract}
RISC-V open-source systems are emerging in deployment scenarios where safety and security are critical.
OpenTitan is an open-source silicon root-of-trust designed to be deployed in a wide range of systems, from high-end to deeply embedded secure environments.
Despite the availability of various cryptographic hardware accelerators that make OpenTitan suitable for offloading cryptographic workloads from the main processor, there has been no accurate and quantitative establishment of the benefits derived from using OpenTitan as a secure accelerator.
This paper addresses this gap by thoroughly analysing strengths and inefficiencies when offloading cryptographic workloads to OpenTitan.
The focus is on three key IPs — HMAC, AES, and OpenTitan Big Number accelerator (OTBN) — which can accelerate four security workloads: Secure Hash Functions, Message Authentication Codes, Symmetric cryptography, and Asymmetric cryptography.
For every workload, we develop a bare-metal driver for the OpenTitan accelerator and analyze its efficiency when computation is offloaded from a RISC-V application core within a System-on-Chip designed for secure Cyber-Physical Systems applications. 
Finally, we assess it against a software implementation on the application core.
The characterization was conducted on a cycle-accurate RTL simulator of the System-on-Chip (SoC).
Our study demonstrates that OpenTitan significantly outperforms software implementations, with speedups ranging from 4.3x to 12.5x.
However, there is potential for even greater gains as the current OpenTitan utilizes a fraction of the accelerator bandwidths, which ranges from 16\% to  61\%, depending on the memory being accessed and the accelerator used.
Our results open the way to the optimization of OpenTitan-based secure platforms, providing design guidelines to unlock the full potential of its accelerators in secure applications.

\end{abstract}



\received{12 January 2024}
\received[revised]{15 February 2024}
\received[accepted]{15 February 2024}

\maketitle

\section{Introduction}
\label{sec:Introduction}

The design of open-hardware RISC-V platforms is increasingly being explored for use in safety and security-sensitive environments like autonomous driving, smart-city infrastructures, and biomedical implantable devices within the research community~\cite{ciani2023cyber, schoenle2018multisensor}. 
Ensuring execution integrity and data protection in deployment in these scenarios is essential to avoid potential financial and human losses. 
Dedicated hardware extensions like cryptographic accelerators, ISA extensions, secure enclaves, or on-chip root-of-trusts are often integrated into these systems to enhance security guarantees and enable efficient secure computation without compromising computational efficiency~\cite{nasahl2021hectorv,lee2020keystone}.

OpenTitan is the first RISC-V, open-source, reference design for the implementation of silicon root-of-trusts. 
It features a set of cryptographic hardware accelerators to compute the most common security primitive required to enforce data confidentiality, integrity and authenticity, such as Secure Hash Functions (SHA-256), Message Authentication Codes (HMAC), block ciphers (AES), and asymmetric encription schemes (OTBN).
Moreover, it gives access to tamper-proof memory storage for keys and cryptographic secrets, and a general-purpose programmable microcontroller to manage the Root-of-Trust operations and to enable the interaction between the OpenTitan and the external world. 
The flexibility of the OpenTitan platform, the possibility to program any security policy in its microcontroller, and the presence of a wide range of cryptographic accelerators made OpenTitan an appealing choice for creating innovative embedded secure RISC-V platforms. 
In fact, current cutting-edge literature works in secure embedded systems utilize OpenTitan to introduce advanced security features like Secure Boot~\cite{wagner2022to}, implementation of sophisticated execution integrity schemes~\cite{parisi2024titancfi} and non-conventional communication channels~\cite{ciani2023cyber}.

In spite of the growing interest in verifying the security of OpenTitan~\cite{meza2023security}, and implementing and deploying new security policies and protocols in OpenTitan~\cite{ciani2023cyber}, there is still a lack of detailed analysis and feasibility assessment regarding its performance as a cryptographic accelerator for security workloads, and datasheet information is not sufficient to accomplish this analysis as it does not take into account the interactions with the memory hierarchy of the system-on-chip. 
Utilizing OpenTitan as a cryptographic accelerator allows leveraging its hardware capabilities to speed up the computation of four critical classes of cryptographic operations, namely secure hash functions, message authentication codes, symmetric and asymmetric cryptography, needed for establishing secure communication channels and ensuring data authenticity and integrity. 
However, two key architectural features threaten the potential performance gains achievable by using OpenTitan as a cryptographic accelerator. 
The OpenTitan secure enclave is physically isolated from the main core of the system-on-chip, which possesses and utilizes the data for encryption. 
As a result, utilizing OpenTitan as a security accelerator requires the secure microcontroller to retrieve the data through the memory hierarchy of the system-on-chip and copy it into the accelerator FIFO, potentially limiting the throughput of accelerators depending on the target system memory hierarchy organization and performance.
Moreover, the original design of OpenTitan lacks DMA or a transparent data movement mechanism. 
This means the responsibility for transferring data from the memory hierarchy to the accelerator FIFO falls entirely on the OpenTitan secure microcontroller. 
As a result, this could further limit accelerator bandwidth due to a lack of efficient memory access mechanisms.
This work is the first attempt in the current state-of-the-art to analyze the effectiveness of the OpenTitan Root-of-Trust as a hardware accelerator for cryptographic workloads. 
This paper presents three primary contributions.
\begin{itemize}
    \item \textbf{Characterization.} We assess the effectiveness of three OpenTitan cryptographic accelerators — HMAC, AES, and OTBN — evaluated on five workloads: SHA-256, HMAC, AES-256, RSA-512, and RSA-1024. 
    These primitives have been selected as central in major libraries and transport layer security protocols, such as TLS and DTLS.
    The assessment has been performed on a cycle-accurate RTL simulator of the OpenTitan design integrated into an SoC design.
    \item \textbf{Performance Analysis.} We assess the performance achiev\-ed by labeling each assembly instruction in the developed cryptographic primitives firmware with its category (such as memory access or control-flow instruction) and purpose (like accelerator initialization or wait for completion). 
    This data, combined with execution time measurements, identifies computational limitations that affect efficiency, providing specific insight into areas for improvement. 
    \item \textbf{Comparison.} We examine the results obtained from the OpenTitan accelerators in comparison to both the maximum theoretical throughput of the hardware accelerators, and a reference software implementation operating on CVA6, a modern RISC-V application-class core that is widely utilized in RISC-V embedded platforms. 
    Our analysis reveals that utilizing OpenTitan results in a performance improvement ranging from 4.3x to 12.5x compared to a pure software implementation.
    However, high latencies in memory access and the lack of transparent data transfer mechanisms restrict system performance to memory bandwidth. 
    This results in utilizing only 16\% to 61\% of the available accelerator bandwidth, which varies based on the level of memory hierarchy being accessed and the type of accelerator used.
\end{itemize}
The rest of the paper is organized as follows. 
Section \ref{sec:background} provides an overview of the OpenTitan architecture and the cryptographic accelerators that we evaluate. 
Additionally, it outlines the integration of OpenTitan as a secure subsystem within the reference SoC~\cite{ciani2023cyber}, and the key architectural features of CVA6~\cite{zaruba2019cost}, the host core used to assess software implementation for security workloads.
Section \ref{sec:methodology} describes the testbed, software, and analysis pipeline we used for benchmarking.
Section \ref{sec:experimental_results} and Section \ref{sec:conclusions} discuss the experimental results, highlighting advantages and criticalities in the proposed benchmarks, and conclude the paper.

\section{Background}
\label{sec:background}

\subsection{OpenTitan Architecture}

OpenTitan~\footnote{\url{https://opentitan.org/}} is a security module available as an open-source system-on-chip design by lowRISC, serving to establish a secure enclave with a hardware Root of Trust (RoT).
The OpenTitan project provides different versions of the same architecture, targeting different applications such as data centers, cloud systems, and embedded systems. 
OpenTitan is designed to protect sensitive data and systems against hardware attacks, tampering, and counterfeiting. 
The specific OpenTitan architecture benchmarked in this study is the \texttt{top\_earlgrey} architecture.
OpenTitan includes a secure microcontroller, on-board private memory, cryptographic hardware accelerators and communication I/O interfaces.
OpenTitan implements a secure boot process that exclusively permits trusted software to execute, ensuring the system's integrity at boot time. 
Additionally, cryptographic acceleration and memory protection mechanisms contribute to swift encryption and secure data storage. 

The hardware architecture centers around the Ibex core, a 32-bit RISC-V CPU designed to support embedded and power-efficient applications~\cite{schiavone2017slow}, optimized for minimal area usage. 
The system-level interconnect is based on the TileLink Uncached Lightweight (TL-UL) protocol. 
The memory domain of OpenTitan consists of a scratchpad SRAM memory and an embedded flash memory, each equipped with separate controllers including Error Correcting Code (ECC) and data-address scrambling to enhance security and reliability. 
Essential cryptographic and scrambling keys are generated by accelerators and stored in a secure, tamper-proof region: the One Time Programmable eFuse memory. 
It incorporates a key manager responsible for managing hardware identities and root keys and protecting critical assets from software threats.
The security domain also includes a hardware life cycle controller, which dynamically enables or disables features based on the device's life cycle state.
OpenTitan features a peripheral subsystem supporting UART, USB, I2C, SPI, and GPIO for communication with the external world. 
Since OpenTitan is designed as a stand-alone SoC, some extensions are required to integrate it into broader SoCs. 
Remarkably, it is necessary to expose to the in-out interface a master port driven by the main TL-UL interconnect, along with a bridging module to convert the TL-UL protocol, adopted by OpenTitan, into the one featured by the host platform.

Dedicated hardware accelerators have been incorporated to handle cryptographic computations efficiently. 
These accelerators are specifically designed to support three essential cryptographic algorithms mandated by the security features targeted by the silicon Root of Trust, namely the Advanced Encryption Standard (AES), Secure Hash Functions (SHA-2 and SHA-3), and Message Authentication Codes (HMAC and KMAC).
Moreover, the OpenTitan Big Number accelerator (OTBN) enables accelerating asymmetric encryption algorithms, such as RSA, which are crucial to key exchange mechanisms and digital signature schemes.

\subsubsection{HMAC and SHA-256}
SHA-256 is a secure hashing function widely utilized to ensure data integrity in applications such as digital signatures and certificate generation. 
This algorithm processes an input message to produce a hash value of fixed size, employing one-way design that prevents reverse engineering or finding different inputs leading to the same hash value. 
HMAC generates a secure digest using cryptographic hash functions, creating a message authentication code which depends on a secret key shared between sender and recipient.
OpenTitan includes an HMAC accelerator that can be programmed to calculate both SHA-256 and HMAC message digests. 
The front-end of the accelerator comprises a memory-mapped FIFO where the software pushes the next 64-byte data block for processing. 
Once the FIFO is filled, the accelerator automatically reads it and initiates an 80-cycle procedure to update its internal state before waiting for the next data block to be processed.
To utilize the HMAC accelerator, the software sets up the accelerator in either SHA-256 or HMAC mode and specifies the message and digest endianness. 
If configured in HMAC mode, the software also loads the secret key into designated control registers. 
It then enters a loop in which it continuously waits for the FIFO to be empty before pushing the next 64-byte data block.

\subsubsection{AES}
The Advanced Encryption Standard (AES) is a symmetric encryption algorithm that operates on a 16-byte block, supporting key lengths of 128, 192, and 256 bits. 
The AES accelerator in the OpenTitan Root-of-Trust optimizes storage utilization through real-time generation of round keys, reducing storage needs and accelerating key-switching activities. 
Upon reset, or when activated by software, the device erases the main key and data registers using pseudo-random data to reduce the possibility of side-channel leakage.
The front-end of the accelerator comprises a 16-byte FIFO where the plaintext is loaded by the software driver. 
When configured in automatic mode, upon loading of the input FIFO, the accelerator immediately encrypts its content and generates the next ciphertext block based on the chosen mode of operation. 
This work focuses on an OpenTitan design that incorporates the default masked AES accelerator implementation and utilizes a CBC mode with a 256-bit key. 
Under these conditions, the encryption of a 16-byte plaintext block takes 72 cycles. 
The operational principles of this accelerator are similar to those of the HMAC accelerator. 
Once control registers are set up for the accelerator and both key and initialization vector are loaded, the software enters the processing loop within which it awaits completion of acceleration operations before retrieving ciphertext from output FIFO and feeding in subsequent plaintext blocks for processing into input FIFO.

\subsubsection{RSA and OTBN}
OpenTitan enhances the performance of mathematical operations in public-key cryptosystems by delegating them to the OTBN co-processor. 
This co-processor is equipped with a tailored processor for broad integer arithmetic, a 32-bit control path, and a 256-bit data path. 
The OTBN offers comprehensive support for control flow with conditional branches, unconditional jumps, and hardware loops. 
In addition, it implements custom instructions to support common functions in public-key cryptosystems, like pseudo-module addition or partial-word multiplication and accumulation.
This work focuses on RSA, a commonly used form of asymmetric encryption in digital signature and key-exchange protocols. 
Its security is based on the challenge of factoring the product of two large prime numbers to undo the exponentiation operations required for message encryption or decryption.
Accelerating asymmetric encryption on the OTBN involves first implementing the specific algorithm using the OTBN instruction set. 
Then, the Ibex secure microcontroller within OpenTitan loads the program binary and data into reserved memories in OTBN. 
Lastly, a command to start computation is activated on OTBN, with OpenTitan waiting for operation completion by checking a control register in the OTBN interface.

\subsection{CVA6 Architecture}

The CVA6 processor \cite{zaruba2019cost} implements the RV64GC instruction set with three privilege levels — Machine, Supervisor, and User — to support Unix-like operating system.
The pipeline spans six stages: PC generation, instruction fetch, instruction decode, issue stage, execute stage, and commit stage. 
The branch prediction mechanism, crucial for mitigating control flow stalls, employs a branch history table, a branch target buffer, and a return address stack. 
The fetch interface employs light pre-decoding to detect branch jumps and accommodate interleaved compressed instructions. 
CVA6's memory management unit (MMU) provides hardware support for address translation, featuring separate and configurable data and instruction translation look-aside buffers (TLBs). 
Exceptions can occur at various pipeline stages and are linked to specific instructions.

\subsection{Reference System-on-chip Architecture}
\begin{figure*}[t!]
    \includegraphics[width=0.85\textwidth]{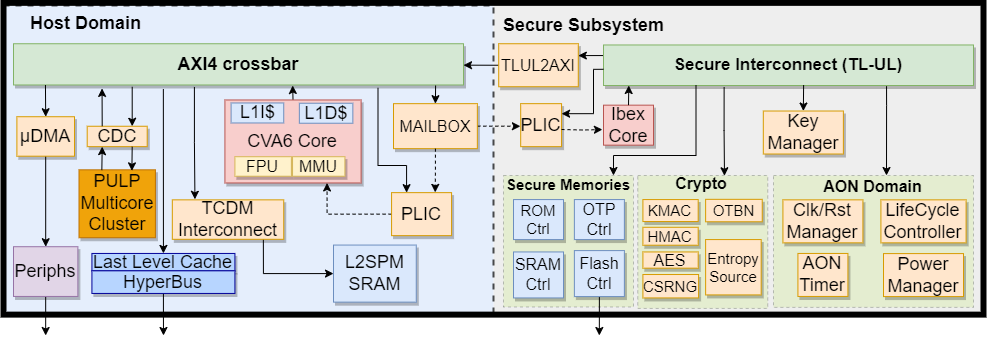}
    \caption{Diagram depicting the architectural integration of OpenTitan as a secure subsystem.}
    \label{fig:soc}
\end{figure*}

To evaluate the effectiveness of OpenTitan Root-of-Trust as an accelerator for cryptographic workloads, we utilized a configuration where OpenTitan is integrated into a larger SoC and can function as a bus master with unrestricted access to all levels of memory within the system-on-chip. 
The SoC used in this research is inspired by the design described in \cite{ciani2023cyber}, shown in Figure~\ref{fig:soc}. 
It incorporates CVA6 as an application processor~\cite{zaruba2019cost}, a PULP-based multicore cluster designed to efficiently handle signal processing and machine learning workloads~\cite{rossi2015pulp}, and a secure subsystem built on the OpenTitan architecture. 
The memory hierarchy of the host comprises private data and instruction cache memories for CVA6 (L1), an embedded scratchpad memory (L2), an hyperbus to access an external HyperRAM (L3) acting as main memory, along with a last-level cache (LLC) that enhances the performance in HyperRAM accesses.

As the application domain of the reference platform employs an AXI4 interconnect while the secure subsystem relies on a TileLink interconnect, a crucial extension to allow OpenTitan to interface with the host is the implementation of a TileLink-AXI4 bridge, named \texttt{TLUL2AXI} in Figure~\ref{fig:soc}. 
This module is connected as a slave to the TLUL interconnect and exposes to the in-out interface of OpenTitan the AXI4 master interfaces, enabling OpenTitan to control the primary interconnect through a dedicated master port, providing visibility on the entire memory map of the SoC.

To preserve the host core from tampering with the content of the secure subsystems, OpenTitan connects to the rest of the SoC through a primary interconnect master port and does not make any slave ports accessible.
Communication between the main processor and the secure component is restricted to exchanging messages via a mailbox system facilitated by interrupts.
The mailbox, in combination with the \texttt{TLUL2AXI} bridge, enables the main processor to assign security-sensitive responsibilities to the secure subsystem. 
This is achieved by inputting into the mailbox registers the location of the buffer containing data for processing, its size and a unique identifier that describes the intended operation.
Additionally, there exists a control register within the mailbox allowing activation of an interrupt signal from OpenTitan to indicate the availability of a new task in the mailbox.
Upon completion of a task, OpenTitan triggers an interrupt to notify CVA6.

\section{Methodology}
\label{sec:methodology}

\begin{table*}
    \centering
    \caption{Meaning of the opcode and semantic labels used to annotate the benchmark execution traces.}
    \label{table:labelling}
    \begin{tabular}{clll}

        \toprule

        Type & IP & Label & Description \\

        \midrule

        \multirow{4}{*}{\rotatebox[origin=c]{90}{OpCode}}
        & \multirow{4}{*}{All} & ALU        & Arithmetic and logic operations between data stored in registers. \\
        &                      & CTL        & Branch, jump, call, and return instructions. \\
        &                      & Memory-RoT & Load/Store instructions accessing the OpenTitan private scratchpad or the accelerators FIFO. \\
        &                      & Memory-RAM & Load/Store instructions accessing the system main memory. \\

        \midrule

        \multirow{7}{*}{\rotatebox[origin=c]{90}{Semantic}}
        & \multirow{4}{*}{HMAC} & Config & Configure the accelerator in SHA-256 or HMAC mode and load the secret key if necessary. \\
        &                       & Digest & Load data from memory and push it into the accelerator FIFO making it ready for processing. \\
        &                       & Wait   & Wait for the accelerator to be ready to accept further data. \\
        &                       & Final  & Pad the message, finalize digest computation and copy the result back to memory. \\
        \cmidrule(lr{0.5em}){2-4}
        & \multirow{3}{*}{AES}  & Config & Initialize the accelerator, set the mode of operations, the key, and the initialization vector. \\
        &                       & Cipher & Push the next data block into the the accelerator and writes the previous block back to memory. \\
        &                       & Wait   & Wait for the accelerator to be ready to accept further data. \\

        \bottomrule

    \end{tabular}
\end{table*}

\subsection{Testbed Preparation}
\label{subsec:testbed_preparation}

Benchmarking the performance of OpenTitan as a cryptographic accelerator necessitates establishing a test environment with two attributes: \textit{SoC Integration} and \textit{Fine-Grained Analysis}.
\begin{itemize}
    \item \textbf{SoC Integration}. Integrating OpenTitan into the same system-on-chip as the host core responsible for cryptographic workloads is crucial. 
    This integration plays a key role in accurately evaluating how the memory hierarchy impacts data access and result delivery during benchmarking.
    \item \textbf{Fine-Grained Analysis}. The integrated SoC deployment should allow us to run the software and precisely record the number of cycles needed for each assembly instruction to complete its execution. 
    This will support accurate analysis of benchmark outcomes and detection of system bottlenecks and overheads.
\end{itemize}
The official OpenTitan releases offer two deployment options. 
One option involves synthesizing the OpenTitan platform on an FPGA, while the alternative workflow synthesizes the hardware for hardware simulation. 
However, neither of these choices align with the \textit{SoC Integration} requirement mentioned earlier. 
The only officially distributed OpenTitan release so far, known as \texttt{top\_earlgrey}, is delivered as a stand-alone system designed to communicate with external devices through embedded systems' off-chip buses like SPI or UART. 
However, using embedded off-chip protocols with limited bandwidth could potentially impact data transfer efficiency to and from an accelerator, thereby affecting the assessment of performance in utilizing the memory hierarchy when benchmarking a hardware accelerator's performance.

When considering the deployment approach, there are two main options for evaluating new system-on-chip architectures: implementing them on an FPGA or simulating the hardware design using an RTL simulator like ModelSim or Verilator. 
Utilizing an FPGA allows for quick prototyping and fast simulation of the SoC's performance on a hardware platform. 
However, it might not offer comprehensive details on the cost of executing each instruction because integrated logic analyzers, a widely used technology for extracting detailed information from FPGA deployments, can only record instruction execution for a limited number of cycles due to restricted buffers. 
Furthermore, most RISC-V system-on-chips, such as OpenTitan, do not include tracing mechanisms capable of tracking instruction execution post-deployment.
On the other hand, software simulators provide an accurate representation of the system under test but are much slower in simulation speed compared to FPGAs. 
In our scenario, the benchmarking software includes a series of bare-metal micro-benchmarks designed to showcase the capabilities of the OpenTitan hardware accelerators. 
Therefore, paying the cost of a slower simulation time is feasible to gain precise control over each instruction's execution and cycle count.

\subsection{Benchmark Implementation}
\label{subsec:benchmark_implementation}

We conducted a series of tests to evaluate the performance achieved by OpenTitan as a security workload accelerator and to compare this performance with that of executing the same workload on the CVA6 host core in software. 
To achieve this, we identified a set of security functions relevant to secure communication channels and data protection use cases. 
For each function, two versions were developed: one for software execution on CVA6, and another programmed for execution on the Ibex secure microcontroller specifically tailored to leverage the capabilities of the OpenTitan hardware accelerator. 
The selected security functions are SHA-256, HMAC, AES-256 (in CBC mode), RSA-512, and RSA-1024. 
In measuring RSA's performance, both data encryption and decryption operations were evaluated due to their differing computational complexities compared with symmetric encryption schemes, like AES, where encryption and decryption are computationally identical.

The software version of the benchmarks is based on BearSSL, an open-source library providing an implementation of all cryptographic algorithms needed for a full SSL/TLS communication stack. 
BearSSL is tailored for deeply embedded systems and its source code is optimized to minimize memory usage. 
The implemented cryptographic algorithms are written in pure C and do not utilize any ISA extension, as CVA6, like most current RISC-V cores intended for deployment in deeply embedded scenarios, does not support any. 
However, when there were multiple implementations available for the same algorithm, we consistently chose the one prioritizing improved performance over binary size or security properties (such as constant-time execution).

When evaluating the performance of hardware accelerators in OpenTitan, the benchmarks are designed for two use-case scenarios: processing input data stored in the system RAM and processing data stored in the OpenTitan private scratchpad. 
The first scenario represents the most common case of the platform host core using OpenTitan as a cryptographic accelerator.
In this scenario, the host core passes to OpenTitan a pointer and a set of metadata presenting the position of the payload in main memory, its size, the operation to be performed, and the address at which the operation result should be stored back.
The second scenario, on the other hand, involves working with sensitive data that had been previously transferred to OpenTitan by the host core and is already located in the OpenTitan private scratchpad memory.  
These latter experiments, in addition to previous ones, demonstrate how memory hierarchy affects OpenTitan's performance as accessing the Root-of-Trust private scratchpad memory is expected to be faster than accessing the main memory outside of the boundaries of the OpenTitan system.  
In both scenarios, the results of cryptographic operations are stored back at the same level in the memory hierarchy from where input data were retrieved, and no analysis is conducted on cases where some data is read from the main memory, processed and stored in the OpenTitan scratchpad or vice versa.
The benchmarks are all written in C and directly program the accelerators, without using the accelerator tests provided with the official OpenTitan software stack. 
While these functions are useful for demonstrating how the hardware accelerators behave, they should not be used in production because they miss some optimization opportunities and perform unnecessary memory accesses that can impact the maximum performance achievable by the hardware accelerators. 
In addition, any situation where software needs to wait for the accelerator to be ready before proceeding (such as when input FIFOs are full or when accelerators need reconfiguration) is specifically managed through polling to avoid additional latencies caused by inefficiencies in interrupt handling and wake-up delays.

\subsection{Benchmark Traces Analysis}
\label{subsec:benchmark_traces_analysis}

The performance analysis of the software running in OpenTitan is performed by combining the information extracted from the execution traces produced by the hardware simulator, which simulates the execution of the benchmarks with a hand-annotated version of the disassembled benchmark software where each assembly instruction is given two labels: a semantic label and an opcode label.
Table \ref{table:labelling} describes the type and meaning of both opcode and semantic labels.
The opcode labels indicate the nature of the operation carried out by the assembly instruction. 
This labeling method emphasizes the balance between computational opcodes and memory accesses, shedding light on which category of instructions is prevalent in the benchmark being analyzed. 
Furthermore, as the execution traces display both the executed instruction and register values stored in its operands, it becomes feasible to differentiate between different types of memory accessed with each memory access, distinguishing between system memory and private OpenTitan scratchpad.

For each benchmark, the semantic label specifies the purpose of executing a given instruction. 
Although the descriptive labels for instructions remain consistent across all benchmarks, semantic labels vary based on the specific accelerator requirements for each run. 
Semantic labeling allows us to identify where most cycles are lost during benchmark execution and determine whether the performance of a particular cryptographic workload is constrained by data input latency into accelerators' FIFOs or by hardware IP bandwidth in producing requested operation results.
This labeling information, combined with the cycle cost of executing each instruction from the execution trace, offers a precise understanding of which operation is most costly when using a specific accelerator. 
Simultaneously, this indirectly assesses the performance of the platform architecture and memory hierarchy.
\section{Experimental Results}
\label{sec:experimental_results}

\subsection{Memory bandwidth characterization}
\label{subsec:preliminary_evaluation}

\begin{table}
    \centering
    \caption{Memory access cost statistics in cycles}
    \label{table:memory_access_cost}
    \begin{tabular}{lrrr}

        \toprule

        \multirow{2}{*}{Memory} & \multicolumn{3}{c}{Latency [cycles]} \\
        \cmidrule(lr{0.5em}){2-4}
                                & Min. & Max. & Avg.                   \\

        \midrule

        RoT &  5.0 & 12.0 &  5.7 \\
        RAM & 23.0 & 27.8 & 23.4 \\

        \bottomrule

    \end{tabular}
\end{table}

\begin{table}
    \centering
    \caption{Memory access bandwidth in cycles/byte}
    \label{table:memory_access_bandwidth}
    \begin{tabular}{lllrrrr}

        \toprule

        \multicolumn{3}{c}{\multirow{2}{*}{Accelerator}} & \multicolumn{2}{c}{HMAC} & \multicolumn{2}{c}{AES} \\
        \cmidrule(lr{0.5em}){4-5}
        \cmidrule(lr{0.5em}){6-7}
                                                     & & & RoT & RAM                & RoT & RAM               \\

        \midrule

        \multicolumn{3}{l}{Block Size [B]} & \multicolumn{2}{c}{64} & \multicolumn{2}{c}{16} \\

        \midrule
        \multirow{6}{*}{Accesses [\#]} & \multirow{2}{*}{Accel.} & Read  & -- & -- &  4 &  4 \\ 
                                       &                         & Write & 16 & 16 &  4 &  4 \\
        \cmidrule(lr{0.5em}){2-7}
                                       & \multirow{2}{*}{RoT}    & Read  & 16 & -- &  4 & -- \\ 
                                       &                         & Write & -- & -- &  4 & -- \\
        \cmidrule(lr{0.5em}){2-7}
                                       & \multirow{2}{*}{RAM}    & Read  & -- & 16 & -- &  4 \\ 
                                       &                         & Write & -- & -- & -- &  4 \\

        \midrule

        \multirow{4}{*}{Cost [cycles]} & \multicolumn{2}{l}{Accel.} &  80 &  80 &  40 &  40 \\
                                       & \multicolumn{2}{l}{RoT}    &  80 &  -- &  40 &  -- \\
                                       & \multicolumn{2}{l}{RAM}    &  -- & 368 &  -- & 184 \\
        \cmidrule(lr{0.5em}){2-7}
                                       & \multicolumn{2}{l}{TOT}    & 160 & 448 &  80 & 224 \\

        \midrule

        \multicolumn{3}{l}{Bandwidth [cycles / B]} &  2.5 &  7.0 &  5.0 & 14.0 \\

        \bottomrule

    \end{tabular}
\end{table}

As a preliminary assessment, we analyze the maximum memory bandwidth achievable by the Ibex secure microcontroller when accessing the system RAM and Root-of-Trust private scratchpad.
Since this system lacks an on-chip DMA, explicit handling of data transfers from memory hierarchy to accelerator FIFO by Ibex is necessary. 
Therefore, evaluating how efficiently Ibex manages data transfers becomes a crucial factor, as it could potentially limit accelerator throughput in memory-bound scenarios. 

Table~\ref{table:memory_access_cost} illustrates the statistics of the memory access cost paid by Ibex when performing a single memory access to the OpenTitan private scratchpad or the system RAM.
This information is derived from collecting memory access costs across various benchmarks, depending on the targeted memory. 
The results indicate that accessing the OpenTitan private scratchpad, or any accelerator FIFO, takes an average of 5.7 cycles, while accessing the system RAM with a LLC hit involves approximately 23.4 cycles. 
The minimum and maximum cycle counts in Table~\ref{table:memory_access_cost} represent the architectural cost of memory accesses, with higher counts occurring after branch or control-flow transfers. 
To assess maximum achievable bandwidth, we consider the minimum memory access cost to gauge platform capability at its peak potential.

Table~\ref{table:memory_access_bandwidth} estimates the minimum $cycles/byte$ achievable by the HMAC and AES accelerators. 
This estimation is based on a calculation of Ibex's memory accesses when collecting input data, which is then multiplied by the memory access cost shown in Table~\ref{table:memory_access_cost}. 
This analysis demonstrates the maximum attainable throughput for these accelerators, since pushing data into accelerator FIFOs is constrained by Ibex's ability to fetch words from memory.
The HMAC accelerator requires fetching from memory and pushing 16 4-byte words per compression loop into its input FIFO, resulting in a minimum bandwidth requirement of 2.5 $cycles/byte$ if data comes from OpenTitan scratchpad or 7.0 $cycles/byte$ if read from system RAM.
A single AES-256 encryption, or decryption, iteration requires loading 4 words from memory, pushing them into the plaintext FIFO, reading back the previously ciphered 16-byte block, and writing it back to memory.
As an effect, AES-256 exhibits greater computational intensity, requiring between 5.0 and 14.0 $cycles/byte$, depending on the memory hierarchy level.

\subsection{OpenTitan performance analysis}

\begin{table*}[t]
    \centering
    \caption{Analysis of OpenTitan HMAC and AES accelerator performance for 4kiB payload}
    \label{tab:opentitan_performance}
    \begin{tabular}{cl*{10}{>{\raggedleft\let\newline\\\arraybackslash}p{.35in}}}

        \toprule

                                                  &
        \multirow{3}{*}{\vspace{-.15in}Operation} & 
        \multicolumn{5}{c}{Instructions [\#]}     & 
        \multicolumn{5}{c}{Cycles [\%]}           \\
            
        \cmidrule(lr{0.5em}){3-7}
        \cmidrule(lr{0.5em}){8-12}

                                           &
                                           & 
        \multirow{2}{*}{\vspace{-.1in}ALU} & 
        \multirow{2}{*}{\vspace{-.1in}CTL} & 
        \multicolumn{2}{c}{Memory}         & 
        \multirow{2}{*}{\vspace{-.1in}TOT} & 
        \multirow{2}{*}{\vspace{-.1in}ALU} & 
        \multirow{2}{*}{\vspace{-.1in}CTL} & 
        \multicolumn{2}{c}{Memory}         & 
        \multirow{2}{*}{\vspace{-.1in}TOT} \\

        \cmidrule(lr{0.5em}){5-6}
        \cmidrule(lr{0.5em}){10-11}

                                &
                                & 
                                &
                                & 
        \multicolumn{1}{c}{RoT} & 
        \multicolumn{1}{c}{RAM} & 
                                & 
                                &
                                & 
        \multicolumn{1}{c}{RoT} & 
        \multicolumn{1}{c}{RAM} & 
                                \\

        \midrule

        \multirow{5}{*}{\rotatebox[origin=c]{90}{SHA-256}}
        & Config &    5 &   -- &    6 &   -- &   11 &  0.02 &    -- &   0.10 &     -- &  0.12 \\
        & Digest &  230 &  210 & 1090 & 1024 & 2554 &  0.92 &  0.68 &  21.06 &  73.24 & 95.90 \\
        & Wait   &   74 &   74 &   74 &   -- &  222 &  0.46 &  0.23 &   2.48 &     -- &  3.17 \\
        & Final  &    4 &   -- &   13 &    8 &   25 &  0.02 &    -- &   0.20 &   0.60 &  0.82 \\
        \cmidrule(lr{0.5em}){2-12}
        & TOT    &  313 &  284 & 1183 & 1032 & 2812 &  1.42 &  0.91 &  23.84 &  73.83 & 32260 \\
        \midrule

        \multirow{5}{*}{\rotatebox[origin=c]{90}{HMAC}}
        & Config &    5 &   -- &   23 &   -- &   28 &   0.02 &     -- &   0.40 &     -- &  0.42 \\
        & Digest &  230 &  210 & 1090 & 1024 & 2554 &   0.91 &   0.67 &  20.89 &  72.64 & 95.11 \\
        & Wait   &   86 &   86 &   86 &   -- &  258 &   0.53 &   0.26 &   2.87 &     -- &  3.66 \\
        & Final  &    4 &   -- &   13 &    8 &   25 &   0.02 &     -- &   0.20 &   0.59 &  0.81 \\
        \cmidrule(lr{0.5em}){2-12}
        & TOT    &  325 &  296 & 1212 & 1032 & 2865 &   1.48 &   0.94 &  24.36 &  73.23 & 32526 \\

        \midrule

        \multirow{4}{*}{\rotatebox[origin=c]{90}{AES-256}}
        & Config &   28 &    9 &   64 &   -- &  101 &  0.05 &  0.01 &  0.54 &    -- &  0.61 \\
        & Cipher & 1281 &  256 & 2306 & 2048 & 5891 &  4.60 &  0.39 & 18.89 & 72.42 & 96.30 \\
        & Wait   &  258 &  257 &  257 &   -- &  772 &  0.77 &  0.39 &  1.93 &    -- &  3.10 \\
        \cmidrule(lr{0.5em}){2-12}
        & TOT    & 1567 &  522 & 2627 & 2048 & 6764 &  5.43 &  0.79 & 21.36 & 72.42 & 66461 \\

        \bottomrule

    \end{tabular}
\end{table*}
\begin{figure*}[t]
    \centering
    \includegraphics[width=0.90\textwidth]{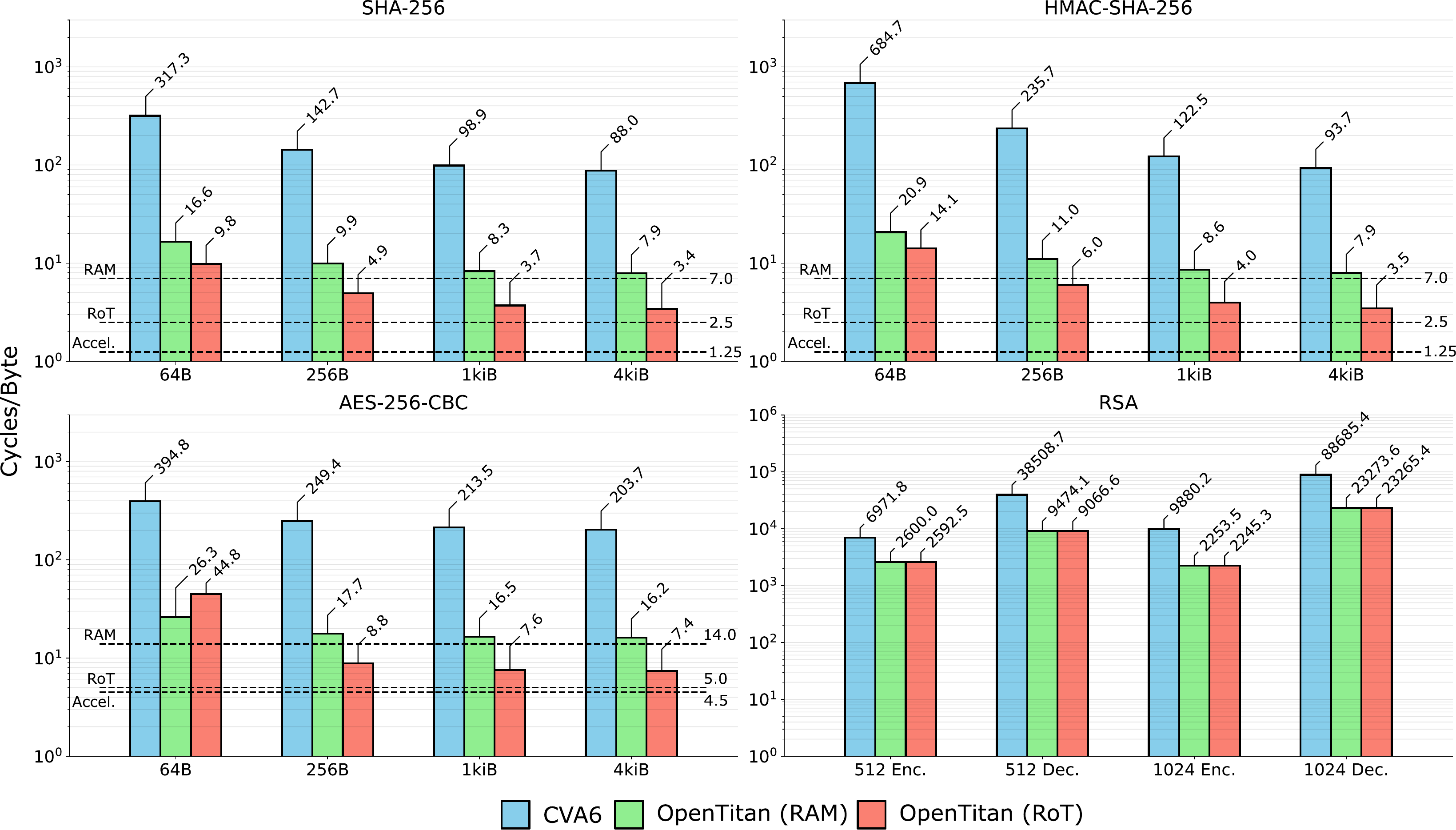}
    \caption{Characterization of the $Cycles/Byte$ reached by the OpenTitan hardware accelerators for different payload sizes.}
    \label{fig:characterization}
\end{figure*}

After evaluating the maximum memory bandwidth, which constrains the maximum achievable throughput by the accelerators, we analyze the effective performance attainable with OpenTitan as a cryptographic workload accelerator for various payload sizes. 
Figure~\ref{fig:characterization} illustrates the $cycles/byte$ needed to process payloads ranging from 64 to 4 kiB using the HMAC and AES IPs, as well as the performance of RSA asymmetric encryption and decryption for key sizes of 512 and 1024 bits. 
The figure shows OpenTitan's achieved performance when loading data from both system RAM and private scratchpad compared to a software implementation on CVA6 core. 
Additionally, they highlight memory bandwidth limits and the theoretical accelerator bandwidth derived from OpenTitan documentation.
SHA-256 and HMAC achieve their highest performance at 7.88 and 7.94 $cycles/byte$, respectively, when data is being loaded from the system RAM. 
When data is loaded from the private scratchpad, these figures improve to 3.41 and 3.47 $cycles/byte$.
In comparison, AES has a lower peak $cycles/byte$ of 16.23 and 7.36.
When it comes to RSA decryption, the most computationally intensive task, a performance of 9\,100 $cycles/byte$ for a 512-bit key, which increases to approximately 23\,300 $cycles/byte$ when utilizing a 1024-bit key.

After evaluating the performance of the OpenTitan accelerators we compared it against CV6 atteined results. 
All benchmarks demonstrate a significant increase in speed compared to running the pure software implementation on the CVA6 host core, even when data is loaded from system RAM. 
For a 4kiB payload, utilizing OpenTitan as a cryptographic accelerator results in an execution speedup of 11.1x for SHA-256, 11.8x for HMAC, and 12.5x for AES-256.
When performing RSA decryption on OpenTitan using 512-bit keys, it is 4.3x faster and with 1024-bit keys it's still significantly quicker at 3.8x faster.
This outcome indicates that although additional software development effort is needed for OpenTitan accelerator drivers and modifications to interface with the OpenTitan Root-of-Trust within the host core, it improves overall computational capabilities when dealing with cryptographic workloads.

On the other hand, while the software version of the cryptographic function is markedly outperformed when loading data from the system RAM, the obtained performances still fall significantly short of saturating the OpenTitan hardware accelerators' maximum theoretical bandwidth. 
Specifically, for HMAC and SHA-256, only 16\% of the theoretical bandwidth is utilized, increasing to 28\% for AES. 
This underscores that despite substantial improvements over a pure software implementation, there remains the potential for substantial speedup in principle.
These results show significant improvements if the input data is stored in the Root-of-Trust private scratchpad. 
In this scenario, due to lower memory access latency as reported in Table~\ref{table:memory_access_cost}, there would be better utilization of hardware accelerators, approximately 37\% for HMAC accelerator and around 61\% for AES would be used out of their respective theoretical bandwidths.

Table~\ref{tab:opentitan_performance} presents a comprehensive summary of the instruction execution counts for SHA-256, HMAC, and AES algorithms, along with the distribution of cycles allocated to different phases when loading a 4kiB payload from system RAM. 
Across all three scenarios examined, the analysis consistently indicates that no less than 93.78\% of cycles are dedicated to memory accesses, with over 72\% of cycles required for retrieving data from system RAM. 
Furthermore, at most 3.66\% of cycles are utilized in waiting for the accelerator to complete its computation before pushing the next data chunk into its FIFO. 
This confirms that while OpenTitan provides significant speedup compared to pure software implementation and is strongly memory-bounded, there exists potential for even greater speedups achievable through OpenTitan.
The experimental results highlight the following major findings:
\begin{itemize}
    \item Memory access cost analysis reveals an average of 5.7 cycles for accessing OpenTitan private scratchpad, contrasting with approximately 23.4 cycles for system RAM.
    \item SHA-256 and HMAC deliver a maximum speeds of 7.88 and 7.94 $cycles/byte$ when data is loaded from system RAM, improving to 3.41 and 3.47 $cycles/byte$ when loaded from the private scratchpad. 
    AES operates at peak $cycles/byte$ of 16.23 and 7.36.
    RSA decryption requires about 9\,100 and 23\,300 $cycles/byte$ for 512-bit and 1024-bit keys. 
    \item OpenTitan hardware accelerators outperforms a software implementation, delivering an 11.1x improvement for SHA-256, 11.8x for HMAC, and 12.5x for AES-256 with a payload of 4 kiB. 
    RSA decryption exhibits a speedup of 4.3x for 512-bit keys and 3.8x for 1024-bit keys.
    \item The system underutilizes the OpenTitan accelerators, with only 16\% utilization for HMAC and SHA-256, and 28\% for AES. 
    When input data is placed in the Root-of-Trust private scratchpad, utilization increases to 37\% and 61\% for HMAC and AES accelerators respectively.
\end{itemize}

\section{Conclusions}
\label{sec:conclusions}

This paper presented a comprehensive evaluation of the OpenTitan Root-of-Trust as a cryptographic accelerator in secure open-hardware system-on-chips.
Our assessment focused five relevant cryptographic workloads — SHA-256, HMAC, AES-256, RSA-512, and RSA-1024 — and compares accelerator performance with data loaded from the private Root-of-Trust scratchpad and from system RAM against pure software implementation on the SoC host core.

The experimental results highlight the potential of using OpenTitan as a cryptographic accelerator, demonstrating a speedup equal to 11.1x for SHA-256, 11.8x for HMAC, and 12.5x for AES-256 with respect to a pure software implementation running on the SoC host core. 
Additionally, RSA decryption on OpenTitan showed a 4.3x increase with 512-bit keys and a 3.8x enhancement with 1024-bit keys. 
The study also emphasizes that the system is strongly memory-bounded, and it utilizes only a fraction of the nominal bandwidth of the hardware accelerators, achieving about 16\% peak bandwidth usage for HMAC and SHA-256, increasing to around 28\% for AES when processing data from system RAM due to high cycle requirements and lack of efficient memory movement mechanisms in the OpenTitan \texttt{top\_earlgrey} considered in this work.

Future works will involve expanding this study to encompass a thorough comparison of the OTBN co-processor and an investigation into possible changes to the architecture aimed at improving the efficiency of the Root-of-Trust and SoC-level memory hierarchy. 
The goal is to maximize the use of hardware accelerators within the Root-of-Trust, leading to more efficient utilization of system resources overall.


\bibliographystyle{ACM-Reference-Format}
\bibliography{main-1author}


\begin{thebibliography}{10}


\ifx \showCODEN    \undefined \def \showCODEN     #1{\unskip}     \fi
\ifx \showDOI      \undefined \def \showDOI       #1{#1}\fi
\ifx \showISBNx    \undefined \def \showISBNx     #1{\unskip}     \fi
\ifx \showISBNxiii \undefined \def \showISBNxiii  #1{\unskip}     \fi
\ifx \showISSN     \undefined \def \showISSN      #1{\unskip}     \fi
\ifx \showLCCN     \undefined \def \showLCCN      #1{\unskip}     \fi
\ifx \shownote     \undefined \def \shownote      #1{#1}          \fi
\ifx \showarticletitle \undefined \def \showarticletitle #1{#1}   \fi
\ifx \showURL      \undefined \def \showURL       {\relax}        \fi
\providecommand\bibfield[2]{#2}
\providecommand\bibinfo[2]{#2}
\providecommand\natexlab[1]{#1}
\providecommand\showeprint[2][]{arXiv:#2}

\bibitem[Ciani et~al\mbox{.}(2023)]%
        {ciani2023cyber}
\bibfield{author}{\bibinfo{person}{Maicol Ciani} {et~al\mbox{.}}} \bibinfo{year}{2023}\natexlab{}.
\newblock \showarticletitle{Cyber Security aboard Micro Aerial Vehicles: An OpenTitan-based Visual Communication Use Case}. In \bibinfo{booktitle}{\emph{2023 IEEE International Symposium on Circuits and Systems (ISCAS)}}. \bibinfo{pages}{1--5}.
\newblock
\urldef\tempurl%
\url{https://doi.org/10.1109/ISCAS46773.2023.10181732}
\showDOI{\tempurl}


\bibitem[Davide~Schiavone et~al\mbox{.}(2017)]%
        {schiavone2017slow}
\bibfield{author}{\bibinfo{person}{Pasquale Davide~Schiavone} {et~al\mbox{.}}} \bibinfo{year}{2017}\natexlab{}.
\newblock \showarticletitle{Slow and steady wins the race? A comparison of ultra-low-power RISC-V cores for Internet-of-Things applications}. In \bibinfo{booktitle}{\emph{2017 27th International Symposium on Power and Timing Modeling, Optimization and Simulation (PATMOS)}}. \bibinfo{pages}{1--8}.
\newblock
\urldef\tempurl%
\url{https://doi.org/10.1109/PATMOS.2017.8106976}
\showDOI{\tempurl}


\bibitem[Lee et~al\mbox{.}(2020)]%
        {lee2020keystone}
\bibfield{author}{\bibinfo{person}{Dayeol Lee} {et~al\mbox{.}}} \bibinfo{year}{2020}\natexlab{}.
\newblock \showarticletitle{Keystone: An Open Framework for Architecting Trusted Execution Environments}. In \bibinfo{booktitle}{\emph{Proceedings of the Fifteenth European Conference on Computer Systems}} (Heraklion, Greece) \emph{(\bibinfo{series}{EuroSys '20})}. \bibinfo{publisher}{Association for Computing Machinery}, \bibinfo{address}{New York, NY, USA}, \bibinfo{numpages}{16}~pages.
\newblock
\urldef\tempurl%
\url{https://doi.org/10.1145/3342195.3387532}
\showDOI{\tempurl}


\bibitem[Meza et~al\mbox{.}(2023)]%
        {meza2023security}
\bibfield{author}{\bibinfo{person}{Andres Meza} {et~al\mbox{.}}} \bibinfo{year}{2023}\natexlab{}.
\newblock \showarticletitle{Security Verification of the OpenTitan Hardware Root of Trust}.
\newblock \bibinfo{journal}{\emph{IEEE Security \& Privacy}} \bibinfo{volume}{21}, \bibinfo{number}{3} (\bibinfo{year}{2023}), \bibinfo{pages}{27--36}.
\newblock
\urldef\tempurl%
\url{https://doi.org/10.1109/MSEC.2023.3251954}
\showDOI{\tempurl}


\bibitem[Nasahl et~al\mbox{.}(2021)]%
        {nasahl2021hectorv}
\bibfield{author}{\bibinfo{person}{Pascal Nasahl} {et~al\mbox{.}}} \bibinfo{year}{2021}\natexlab{}.
\newblock \showarticletitle{HECTOR-V: A Heterogeneous CPU Architecture for a Secure RISC-V Execution Environment}. In \bibinfo{booktitle}{\emph{Proceedings of the 2021 ACM Asia Conference on Computer and Communications Security}} (Virtual Event, Hong Kong) \emph{(\bibinfo{series}{ASIA CCS '21})}. \bibinfo{publisher}{Association for Computing Machinery}, \bibinfo{address}{New York, NY, USA}, \bibinfo{pages}{187–199}.
\newblock
\urldef\tempurl%
\url{https://doi.org/10.1145/3433210.3453112}
\showDOI{\tempurl}


\bibitem[Parisi et~al\mbox{.}(2024)]%
        {parisi2024titancfi}
\bibfield{author}{\bibinfo{person}{Emanuele Parisi} {et~al\mbox{.}}} \bibinfo{year}{2024}\natexlab{}.
\newblock \bibinfo{title}{TitanCFI: Toward Enforcing Control-Flow Integrity in the Root-of-Trust}.
\newblock
\newblock
\showeprint[arxiv]{2401.02567}~[cs.CR]


\bibitem[Rossi et~al\mbox{.}(2015)]%
        {rossi2015pulp}
\bibfield{author}{\bibinfo{person}{Davide Rossi} {et~al\mbox{.}}} \bibinfo{year}{2015}\natexlab{}.
\newblock \showarticletitle{PULP: A parallel ultra low power platform for next generation IoT applications}. In \bibinfo{booktitle}{\emph{2015 IEEE Hot Chips 27 Symposium (HCS)}}. \bibinfo{pages}{1--39}.
\newblock
\urldef\tempurl%
\url{https://doi.org/10.1109/HOTCHIPS.2015.7477325}
\showDOI{\tempurl}


\bibitem[Schönle et~al\mbox{.}(2018)]%
        {schoenle2018multisensor}
\bibfield{author}{\bibinfo{person}{Philipp Schönle} {et~al\mbox{.}}} \bibinfo{year}{2018}\natexlab{}.
\newblock \showarticletitle{A Multi-Sensor and Parallel Processing SoC for Miniaturized Medical Instrumentation}.
\newblock \bibinfo{journal}{\emph{IEEE Journal of Solid-State Circuits}} \bibinfo{volume}{53}, \bibinfo{number}{7} (\bibinfo{year}{2018}), \bibinfo{pages}{2076--2087}.
\newblock
\urldef\tempurl%
\url{https://doi.org/10.1109/JSSC.2018.2815653}
\showDOI{\tempurl}


\bibitem[Wagner et~al\mbox{.}(2022)]%
        {wagner2022to}
\bibfield{author}{\bibinfo{person}{Alexander Wagner} {et~al\mbox{.}}} \bibinfo{year}{2022}\natexlab{}.
\newblock \showarticletitle{To Be, or Not to Be Stateful: Post-Quantum Secure Boot Using Hash-Based Signatures}. In \bibinfo{booktitle}{\emph{Proceedings of the 2022 Workshop on Attacks and Solutions in Hardware Security}} (Los Angeles, CA, USA) \emph{(\bibinfo{series}{ASHES'22})}. \bibinfo{publisher}{Association for Computing Machinery}, \bibinfo{address}{New York, NY, USA}, \bibinfo{pages}{85–94}.
\newblock
\urldef\tempurl%
\url{https://doi.org/10.1145/3560834.3563831}
\showDOI{\tempurl}


\bibitem[Zaruba et~al\mbox{.}(2019)]%
        {zaruba2019cost}
\bibfield{author}{\bibinfo{person}{Florian Zaruba} {et~al\mbox{.}}} \bibinfo{year}{2019}\natexlab{}.
\newblock \showarticletitle{The Cost of Application-Class Processing: Energy and Performance Analysis of a Linux-Ready 1.7-GHz 64-Bit RISC-V Core in 22-nm FDSOI Technology}.
\newblock \bibinfo{journal}{\emph{IEEE Transactions on Very Large Scale Integration (VLSI) Systems}} \bibinfo{volume}{27}, \bibinfo{number}{11} (\bibinfo{year}{2019}), \bibinfo{pages}{2629--2640}.
\newblock
\urldef\tempurl%
\url{https://doi.org/10.1109/TVLSI.2019.2926114}
\showDOI{\tempurl}


\end{thebibliography}

\end{document}